\documentstyle[eqsecnum,epsfig, aps,prd]{revtex} 
\tighten
\let\jnfont=\rm
\def\NPB#1,{{\jnfont Nucl.\ Phys.\ }{\bf B#1},}
\def\PLB#1,{{\jnfont Phys.\ Lett.\ B }{\bf #1},}
\def\PRD#1,{{\jnfont Phys.\ Rev.\ D }{\bf #1},}
\def\PRL#1,{{\jnfont Phys.\ Rev.\ Lett.\ }{\bf #1},}

\def\lsim{\mathrel{\mathpalette\oversim<}}
\def\gsim{\mathrel{\mathpalette\oversim>}}
\def\oversim#1#2{\lower0.5ex\vbox{\baselineskip0pt\lineskip0pt
                 \lineskiplimit0pt\everycr{}\tabskip0pt
                 \halign{$\mathsurround0pt #1\hfil##\hfil$\crcr #2\crcr\sim\crcr}}}
\begin{document}
\draft

\preprint{
\vbox{\hbox{\bf TU-721}
      \hbox{\bf hep-ph/0405186} }}

\title{ SuperWIMP dark matter scenario in light of WMAP}

\author{\ \\[2mm] Fei Wang$^1$, Jin Min Yang$^{1,2}$} 

\address{ \ \\[2mm]
 $^1$ {\it Institute of Theoretical Physics, Academia Sinica, 
           Beijing 100080, China}\\
 $^2$ {\it Department of Physics, Tohoku University,
               Aoba-ku, Sendai 980, Japan} }

\maketitle

\begin{abstract}
The heavy gravitino in the minimal supergravity (mSUGRA) models is likely to be the 
lightest supersymmetric particle (LSP).
Produced from the late decays of the metastable Weakly Interacting Massive Particles (WIMPs)
such as the lightest neutralinos, the stable gravitinos can be plausible candidates for the cold dark matter 
in the universe.  Such gravitino dark matter can naturally evade the current detection experiments due to its 
superweak couplings.  However, this scenario must be subjected to the constraints from the Big Bang nucleosynthesis 
(BBN) predictions for light element abundances as well as the Wilkinson Microwave Anisotropy Probe (WMAP) data
for the relic density. Assuming the popular case in which the lightest neutralino is the next-to-lightest 
supersymmetric particle (NLSP), we find that requiring BBN predictions for light element abundances to agree 
with the WMAP data  
can impose upper and lower mass bounds on  both the gravitino LSP and the neutralino NLSP.
A scan over the mSUGRA parameter space, subjected to the BBN constraints, the WMAP data and the $b\to s \gamma$ 
bounds, shows that the low $\tan\beta$ ($\lsim$ 40) region as well as the region accessible at CERN Large 
Hadron Collider (LHC) will be severely constrained.  Such stringent constraints on the parameter space
might be instructive for testing this scenario in future collider experiments. 
 
\end{abstract}


\section{Introduction}
The nature of the dark matter is one of the mysteries in today's physical science.
It has been intensively explored both theoretically and experimentally.  
Studies showed that the cosmic dark matter is plausibly  composed of non-baryonic 
Weakly Interacting Massive Particles (WIMPs)  \cite{review}.
While the Standard Model of particle physics
cannot provide a candidate for the dark matter WIMP, the popular supersymmetric theory 
with R-parity conservation can provide a good candidate, i.e.,  the lightest supersymmetric particle (LSP). 
So far the widely studied scenario is that the lightest neutralino is assumed to be the LSP.
However, despite of the overwhelming popularity of this scenario, 
other possibilities should not be ignored due to the following reasons.
On the one hand, the success of such neutralino dark matter scenario may be spoiled by the problems 
caused by gravitino in the reheating era \cite{spoil}. On the other hand, the neutralino dark matter 
scenario has not yet been confirmed by current experiments\cite{sudan}. 

One possible scenario other than neutralino LSP is that the gravitino is assumed to be the LSP. 
Such gravitino LSPs can form warm or cold cosmic dark matter, depending on the gravitino  mass:  
\begin{itemize}
\item[(i)] In some low-energy SUSY breaking models, like the gauge mediated SUSY breaking (GMSB) models,
           the gravitino can be as light as KeV, much lighter than other supersymmetric particles. It can thus form 
           the warm dark matter. Note that the recent WMAP data imposed severe constraints on the dark matter 
           type. As analyzed in \cite{warm}, while a very tiny component of dark matter can be the hot neutrinos,
           the warm dark matter is ruled out due to the detected early re-ionization of the universe 
           at a redshift $z \approx 0.20$.
           Therefore, the scenario of warm dark matter gravitino is not favored by recent observation.
\item[(ii)] In the popular mSUGRA models, the gravitino mass is unspecified and only known to be of the
            weak-scale. Such heavy gravitino is possibly the LSP and can form the cold dark matter in
            the universe. In contrast to the highly constrained scenarios, in which the gravitino is produced as
            a thermal relic \cite{relic} or produced during reheating \cite{reheat}, a new scenario, assuming 
            the gravitino to be produced from the late decays of the thermal relic WIMPs, was recently proposed 
            in \cite{feng}. 
\end{itemize}
In this article we focus our attention on this new  gravitino dark matter scenario. 
Since the heavy gravitinos couple gravitionally, they are naturally the so-called  Superweakly 
Interacting Massive Particles (SuperWIMPs).  
As a plausible candidate for the cold dark matter in the universe \cite{feng},  
the gravitino SuperWIMP can naturally evade the current dark matter detection experiments due to its superweak 
couplings.  However, this scenario must be subjected to the constraints from the Big Bang nucleosynthesis (BBN)  
as well as the WMAP data \cite{wmap}:  
\begin{itemize}
\item The late decays of WIMPs (like neutralinos) into gravitino SuperWIMPs will release 
      electromagnetic (EM) and hadronic energy. Such energy release will alter the BBN predictions for 
      light element abundances \cite{ellis,hadronic}. Requiring the resulted predictions for 
      light element abundances to agree with the measured values will impose strong constraints 
      on the gravitino dark matter scenario.
\item WMAP precisely measured many quantities, especially the total matter density and the baryon density, 
      \begin{eqnarray} 
      \Omega_m h^2  =  0.135_{-0.009}^{+0.008} \ , ~~~
      \Omega_b h^2  =  0.0224_{-0.009}^{+0.009} \  . 
      \end{eqnarray} 
      From such results we can deduce the 2$\sigma$ range for the cold dark matter density
      \begin{eqnarray}
      \label{density} 
      \Omega_{CDM} h^2 = 0.1126_{-0.0181}^{+0.0161}\ , 
      \end{eqnarray} 
      which is dramatically more accurate than previous results and agrees quite well with other approaches.
      Such precise measurements will impose strong constraints on the parameter space of 
      gravitino dark matter scenario.     
\end{itemize} 
The aim of this article is to examine these constraints on the parameter space of mSUGRA in this 
new gravitino dark matter scenario.  Assuming the popular case that the lightest neutralino 
is the NLSP, we will examine the constraints on the mSUGRA parameter space from the BBN light element 
abundances, the WMAP data of relic density as well as the $b\to s \gamma$ branching ratio data. 
 
\section{Constraints on the parameter space}
\label{sec2}  
In the new scenario \cite{feng}, the NLSPs freeze out with a thermal relic density 
$\Omega_{NLSP}$ and then decay to the gravitino at time $10^4 \sim 10^8 s$.  
Thus the relic density of gravitino dark matter is obtained by 
\begin{equation}
 \Omega_{\tilde G}=\frac{m_{\tilde{G}}}{M_{NLSP}}\Omega_{NLSP} .
\end{equation}
The late decays of the NLSPs will release energy which will alter the light element abundances.
In fact, the later injection of high energy photon with the stopping energy
  inversely proportional to the temperature by scattering off the background
  photon will dissociate the existing light elements.  
If carefully chosen, such injecting EM energy can destruct the light element 
abundances to proper values.
It is well known that the BBN predictions for light element abundances are quite successful 
for most light elements. Yet the BBN predictions for $^{4}{\rm  He}$ and $^{7}{\rm Li}$
(especially $^{7}{\rm Li}$)  
seemingly do not agree with the WMAP data \cite{lecture}.  Requiring such discrepancy 
be settled by the energy released from the late-decaying NLSPs and, at the same time, 
requiring such energy release not to spoil the successful BBN predictions for other light 
elements, the constraints on the NLSP lifetime (in seconds) and EM energy release 
can be obtained \cite{ellis}:
\begin{eqnarray}
\label{range1}
&& 1.5 \times 10^6 s   \leq \tau \leq 4 \times 10^6 s, \\
\label{range2}
&& 0.8 \times 10^{-9} {\rm GeV} \leq \zeta_{EM} \leq 1.5 \times 10^{-9} {\rm GeV}, 
\end{eqnarray}
where 
\begin{eqnarray}
\zeta_{EM}=\epsilon_{\rm EM} B_{\rm EM} Y_{\rm NLSP}
\end{eqnarray}
is the emitted EM energy density, with $\epsilon_{\rm EM}$ being the initial EM energy release from the decay,
$B_{\rm EM}$ being the branching fraction of the decay into EM components, and 
$Y_{\rm NLSP}=n_{\rm NLSP}/n_{\gamma}$ being the NLSP number density normalized to the 
BG photon number density.   
In the derivation of the above bounds, the ratio $\eta\equiv n_{B}/n_{\gamma}$ is fixed to be 
$6 \times 10^{-10}$, and the bounds on the primordial abundances of light elements are taken to be \cite{ellis} 
\begin{eqnarray}
\label{dh}
&& 1.3\times 10^{-5} \ < \ {\rm D}/{\rm H} \ < \  5.3\times 10^{-5} \ , \\
&& 0.227 \ < \ Y_p \ < \ 0.249 \ , \\
&& 9.0\times 10^{-11} \ < \  ^7{\rm Li}/{\rm H} \ < \  2.8 \times 10^{-10} \ , \\
&& ^6{\rm Li} /^7{\rm Li} \ \lsim  \ 0.07  \ ,\\
&& ^6{\rm Li} /{\rm H} \ \lsim  \ 2 \times 10^{-11} \ , 
\end{eqnarray}
where $Y_p$ denotes $^4$He abundance. 

\begin{figure}[tb] 
\centerline{\psfig{file=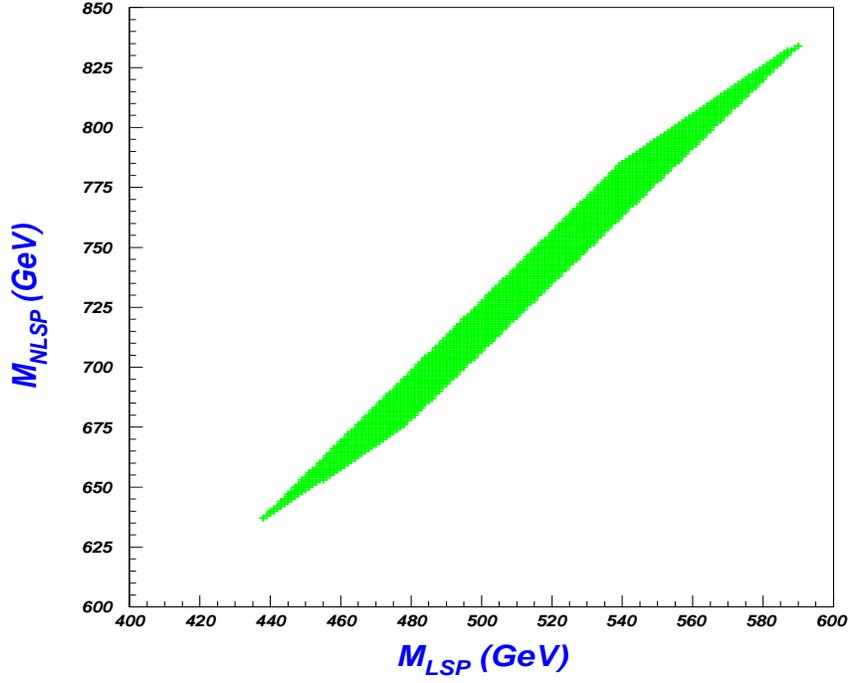,width=12cm,height=10cm,angle=0} }
\vspace*{0.5cm}
\caption{The region (shaded) allowed by BBN to account for $^{7}Li$ and  $^{4}He$ abundance
         in the scenario of gravitino LSP and neutralino NLSP. }
\label{fig1}
\end{figure}

The dominant decay of neutralino ($\tilde\chi$) NLSP into gravitino ($\tilde G$) LSP 
is through $\tilde\chi \to \gamma \tilde{G}$ with a rate given by 
\begin{equation}
\label{rate}
 \Gamma(\tilde\chi \to \gamma \tilde{G})=\frac{|N_{11} \cos\theta_W
  + N_{12} \sin\theta_W|^2}{48\pi M_*^2} {\frac{m_{\tilde\chi}^5}{m_{\tilde G }^2}}
  \left[1-{\frac{m_{\tilde G }^2}{m_{\tilde\chi}^2}}\right]^3 
  \left[1+3\frac{m_{\tilde G }^2}{m_{\tilde\chi}^2}\right], 
\end{equation}
where $m_{\tilde G }$ is the gravitino LSP mass, $m_{\tilde\chi}$ is the neutralino NLSP mass,
$M_*=1/\sqrt{8\pi G_N}\simeq 2.4\times 10^{18}$ GeV is the reduced Plank scale, 
$\theta_W$ is the weak mixing angle, and $N_{ij}$ denotes the matrix element projecting 
the $i$-th neutralino into Bino ($j=1$), Wino ($j=2$) and Higgsinos ($j=3,4$). 
For such a decay we have 
\begin{eqnarray}
&& B_{\rm EM}=1 \ , \\
&& \epsilon_{\rm EM}=\frac{m_{\tilde\chi}^2-m_{\tilde G }^2}{2m_{\tilde\chi}} \ .
\end{eqnarray}
We would like to make some clarifications about our numerical calculations:
\begin{itemize} 
\item[(1)] In the calculation of the thermal relic density of the NLSPs, we considered the 
           general mixing of neutralinos and used the package Microomega \cite{omega},
           which includes all tree level contributions to 
           the scattering amplitudes \footnote{In the present version of Microomega, the package 
           FeynHiggs\cite{feynhiggs} is used to calculate the Higgs masses and Hdecay \cite{hdecay}
           is used to include important QCD corrections to the Higgs decays}. 
\item[(2)] In the scan over the mSUGRA parameter space, we used SuSpect2.0 \cite{suspect} to 
            obtain the sparticle masses in mSUGRA models, which includes one-loop corrections 
            to sparticle masses and two-loop corrections to Higgs masses.  
\item[(3)] When deriving the constraints on the mSUGRA parameter space,  
           we consider the BBN constraints in Eqs.(\ref{range1},\ref{range2}), the WMAP data of relic dark 
           matter density in Eq.(\ref{density}) as well as the $b\to s \gamma$ bound \cite{bsg}:
\begin{eqnarray} 
     2.16 \times 10^{-4} < BF(b\to s \gamma) < 4.34 \times 10^{-4} \ .
\end{eqnarray}
 Furthermore, we fixed the common trilinear coupling $A_{0}=0$ since our results are not sensitive to it.
\item[(4)] In our calculation we also considered the muon anomalous magnetic moment $\alpha_\mu$.
           Since so far much theoretical uncertainty exists in  $\alpha_\mu$ predictions\footnote{
           For example, one group \cite{hagiwara} gives $11.5\leq \Delta \alpha_\mu \times 10^{10} \leq60.7$ or 
           $-16.7 \leq \Delta \alpha_\mu \times 10^{10} \leq 49.1$, depending on the calculation
           approaches. Another group \cite{davier} gives $\delta \alpha_{\mu} \times 10^{10} =12.4_{-8.3}^{+8.3}$ 
           by using $\tau$ decay data to determine vacuum polarization.},  
        we did not use $\alpha_\mu$ to set constraints and, instead, we only gave the range of the mSUGRA 
        contributions corresponding to each case. This will be useful when $\alpha_\mu$ uncertainty is further
        reduced in the future.   
\end{itemize}
The bounds on the masses of the neutralino NLSP and gravitino LSP are shown in Fig.\ref{fig1}.
One sees that BBN results give rather stringent upper and lower bounds
\begin{equation}
430 {\rm~GeV} \leq M_{LSP} \leq 600 {\rm~GeV}, ~~630 {\rm~GeV}  \leq M_{NLSP}\leq 840 {\rm~GeV}. 
\end{equation}
We found that the neutralino NLSP is quite bino-like ($N_{11}\gsim 0.99$) in the allowed regions in Fig.\ref{fig1}  

\begin{figure}
\vspace*{-1cm}
\centerline{\psfig{file=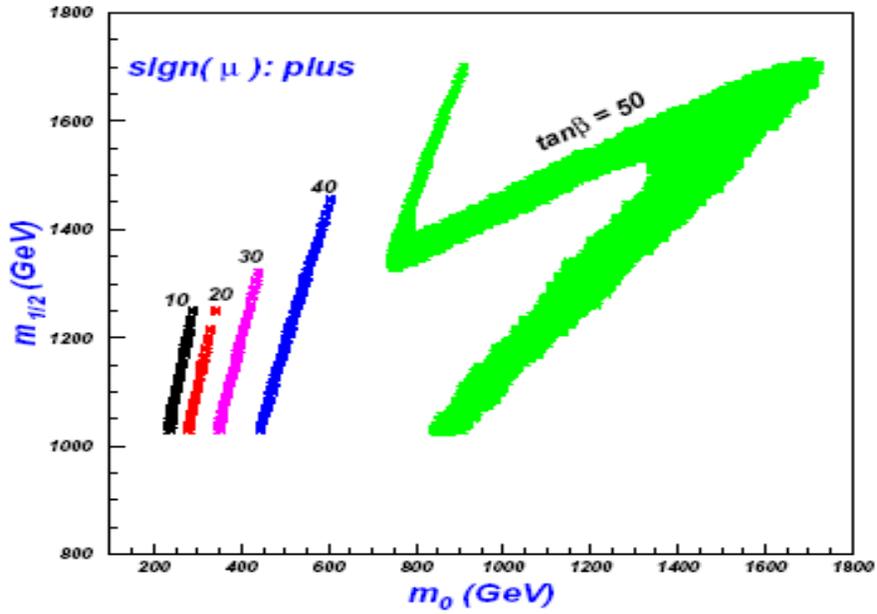,width=13cm,height=12cm,angle=0} }
\vspace*{-2cm}
\caption{The regions (shaded) allowed by the WMAP data, BBN constraints as well as  $b \to s \gamma$ data.
         From left to right corresponds to $\tan\beta=10, 20, 30, 40, 50$, respectively. The sign of $\mu$ is
         assumed to be plus. }
\label{fig2}
\end{figure}

\begin{figure} 
\centerline{\psfig{file=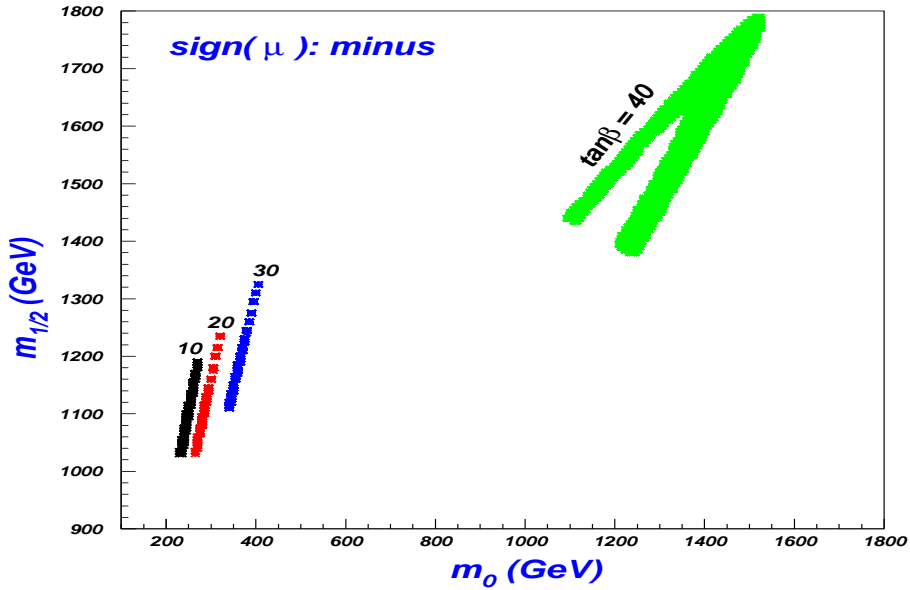,width=13cm,height=8cm,angle=0} }
\caption{Same as Fig.2, but for minus sign of $\mu$. }
\label{fig3}
\end{figure}

The allowed regions in the plane of $m_0$ versus $m_{1/2}$ are shown in Fig.\ref{fig2} for plus sign 
of $\mu$ and Fig.\ref{fig3} for minus sign of $\mu$. From  Fig.\ref{fig2}  we see that 
for plus sign of $\mu$ much of the parameter space with $\tan\beta \lsim 40$ is ruled out. 
      For example, for $\tan\beta=10$, only a narrow strip survives, i.e.,
      225 GeV $\lsim m_0 \lsim $ 300 GeV and  1020 GeV $\lsim m_{1/2} \lsim$ 1270 GeV
      (the corresponding contribution to $\alpha_\mu$ is  
       $1.38\lsim \Delta \alpha_\mu \times 10^{10} \lsim 2.05$).     
      When  $\tan\beta$ increase to $50$, the allowed region gets quite large, i.e.,  
     740 GeV $\lsim m_{0} \lsim$ 1730 GeV and 1060 GeV $\lsim m_{1/2} \lsim$ 1720 GeV 
     (the corresponding contribution to $\alpha_\mu$ is $2.18 \lsim \Delta \alpha_\mu \times10^{10}\lsim 6.32$).

For minus sign of $\mu$,  similar results are obtained.
      We see from Fig.\ref{fig3}  that for $\tan\beta \lsim 30$, the constraints are quite stringent.
      For example, for $\tan\beta=10$ the constraints are 220 GeV $\lsim m_{0} \lsim$ 305 GeV 
      and 1020 GeV $\lsim m_{1/2} \lsim$ 1280 GeV (the corresponding contribution to $\alpha_\mu$ is 
      $-6.07\lsim \Delta \alpha_\mu \times10^{10} \lsim -3.7$).
      When $\tan\beta$ increases to $40$, the constraints are weakened to  
     1060 GeV $\lsim m_{0} \lsim$ 1530 GeV, 1290 GeV $\lsim m_{1/2} \lsim$ 1795 GeV 
    (the corresponding contribution to $\alpha_\mu$ is $-5.45\lsim  \Delta \alpha_\mu \times10^{10} \lsim -1.90$).  

It is interesting to note that if $\Delta \alpha_\mu$ is required to be positive (negative), then 
the allowed regions in $\mu < 0$ ($\mu > 0$)  case will be completely ruled out.
Therefore, a further scrutiny of $\alpha_\mu$ will provide meaningful constraints on
the parameter space of mSUGRA.

It should be pointed out that the above stringent constraints in Figs. 2 and 3 were obtained under 
the requirement that the gravitinos from the late decays of the NLSP make up all the dark matter. 
If other particles like axions are also a component of dark matter, then the constraints will be changed.
Moreover, if the relic gravitino dark matter are partly produced during reheating
\footnote{The gravitinos can also be produced as a thermal relic at very early times. However,
 in the context of inflation, the universe inflated between that early time and now, which would
 dilute any gravitino thermal relic density.},  the constraints will also be changed.
In order to forbid gravitino (with mass range in Eq.(16)) production during reheating, 
the reheating temperature $T_R$ must be lower than a certain value. Following the analyses 
in \cite{reheat},  we evaluated such a limit and found $T_R \lsim 10^9$ GeV.

Note that
in our analysis we only considered the consequence of EM energy release from the late decays of WIMPs  
and required such  EM energy release  to
settle the discrepancy between the BBN predictions and the observed values for light element
abundances.  However, as can be seen in the allowed $M_{NLSP}/M_{LSP}$ region of Fig. 1, 
the decay $NLSP \to \tilde G  + Z$ is also possible and thus will cause hadronic energy release 
from the subsequent $Z$ decays. Such hadronic energy release can be quite dangerous since it can
alter the BBN predictions for the light element abundances.
 
In Ref.\cite{su} the authors studied the bounds from the hadronic energy release 
and found that such hadronic bounds may be stringent\footnote{In their study they used the bounds $
2.4 \times 10^{-5} < D/H < 3.2\times 10^{-5}$, which are more aggressive than the bounds of 
Eq.(\ref{dh}) used by us to obtain the best fit region.} due to the overproduction of $D$.
However, as pointed out in  Ref.\cite{su}, for the best fit region with lifetime $\tau$ 
between $10^6 s$ and  $10^7 s$, the $D$ overproduction from hadronic energy 
is possibly canceled by the overdestruction from EM energy and thus the hadronic bounds become 
less stringent and subject to large uncertainty. In such a region the EM energy effects are dominant \cite{su}. 
Since in our study we focused on the best fit region, we did not consider the hadronic effects.
   
\section{Conclusions}

We examined the constraints on the newly proposed dark matter scenario, 
in which the gravitino is assumed to be the LSP and produced from the late decays of 
metastable NLSP. Although such gravitino dark matter can naturally evade the current detection 
experiments due to its superweak couplings, we found, however, that this scenario is subjected to 
stringent constraints from the BBN predictions for light element abundances 
as well as the WMAP data for the relic density. 
Assuming the popular case that the lightest neutralino is the NLSP in mSUGRA models, 
we found that the low $\tan\beta$ ($\lsim$ 40) region as well as the region accessible at the 
LHC are severely constrained.  

The popular mSUGRA models will be explored in future colliders like the LHC.
For this purpose, it is important to know which part of the parameter space is viable and thus
should be primarily explored. In this regard, the stringent cosmological constraints on the mSUGRA parameter 
space obtained in this work will be useful. Especially, if the ongoing and planned dark matter detection 
experiments fail to find any dark matter signal, it will imply that the dark matter interactions are too weak 
and thus the gravitino dark matter scenario will be favored. Then, to test this scenario at colliders, 
the stringent cosmological constraints on the parameter space will be quite instructive. 
This would serve as a good example that the studies in cosmology and astrophysics 
can shed some light on particle collider physics. 

On the other hand, the LHC could explore mSUGRA parameter space up to $m_{1/2}\sim 1400$ GeV (700 GeV) for 
small (large) values of $m_0$, assuming 100 $fb^{-1}$ of integrated luminosity \cite{LHC}.
If the LHC results finally restrain the parameter space to one of the regions obtained in this work, 
then it implies that the gravitino dark matter scenario is favored. In this sence, the studies
in particle physics can provide some insights in the understanding of dark matter in cosmology. 

We address that our study in this work is just illustrative instead of exhaustive. 
We assumed the popular case that the lightest neutralino is the NLSP in mSUGRA models.
Actually, other super particles, like tau-slepton, are also likely to be the NLSP in mSUGRA models.
If tau-slepton is assumed to be the NLSP, there are some theoretical uncertainties 
in its decay modes and the corresponding energy release.  

Note added: While we are preparing this manuscript, some other preprints \cite{su} appeared,
where the constraints on the gravitino dark matter scenario are studied.
We found that the studies in \cite{su} are quite exhaustively, where the scenarios of 
neutralino NLSP, stau NLSP as well as sneutrino NLSP are all considered. 
Compared with the studies in \cite{su}, the characteristic of our study is that we 
performed a scan over the mSUGRA parameter space and presented the allowed regions 
in terms of original mSUGRA parameters. In addition, the BBN constraints on the
EM energy release are more stringent in our study since we required such energy
release settle the discrepancy between the BBN prediction and the WMAP data for 
$^{7}{\rm Li}$ abundance.      

\section*{Acknowlegents}
We thank Shufang Su for useful comments and Yuqi Lee for debugging the linking code between 
Microomega and SuSpect.  We also thank Junjie Cao and Guangping Gao for helpful discussions.   
This work is supported in part by National Natural Science Foundation of China (NNSFC)
and by a visiting program (No. L03517) of Japan Society for the Promotion of Science.


\begin{thebibliography}{99}
\bibitem{review}  For a recent review, see, e.g., C. Munoz, hep-ph/0309346.
\bibitem{spoil}    S. Weinberg, \PRL48, 1303 (1982);
                   E. Holtmann, M. Kawasaki, K. Kohri,  T. Moroi, \PRD60, 023506 (1999);
                   M. Kawasaki, K. Kohri, T. Moroi, \PRD63, 103502 (2001).
\bibitem{sudan}    D. S. Akerib {\it et al.}, astro-ph/0405033  
\bibitem{warm}     A. B. Lahanas, D. V. Nanopoulos, \PLB568, 55 (2003).
\bibitem{relic}    H. Pagels and J. R. Primack, \PRL48, 223 (1982).
\bibitem{reheat}   M. Bolz, A. Brandenburg and W. Buchmuller, \NPB606, 518 (2001);
                   T. Moroi, H. Murayama and M. Yamaguchi, \PLB303, 289 (1993).
\bibitem{feng}     J. L. Feng, A. Rajaraman, F. Takayama, \PRL91, 011302 (2003);
                           \PRD68, 063504 (2003);  J. L. Feng, hep-ph/0308201.
\bibitem{wmap}     WMAP Collaboration, Astrophys. J. Suppl. {\bf 148}, 1 (2003); 
                   {\bf 148}, 175 (2003); {\bf 248}, 195 (2003).  
\bibitem{ellis}     R. H. Cyburt, J. Ellis, B. D. Fields, K. A. Olive, \PRD67, 103521 (2003).
\bibitem{hadronic}  M. H. Reno and D. Seckel, \PRD37, 3441 (1988);
                    K. Kohri, \PRD64, 043515 (2001);
                    S. Dimopoulos, R. Esmailzadeh, L. J. Hall, G. D. Starkman, \NPB311, 699 (1989);
                    M. Kawasaki, K. Kohri, T. Moroi, astro-ph/0402490. 
\bibitem{lecture}   B. Fields, {\it Primordial  Nucleosythesis in light of WMAP},  
                    CFA colloquium, University of Illinois.
\bibitem{stau}      W.  Buchmuller, K. Hamaguchi, M. Ratz, T.  Yanagida,  hep-ph/0403203.   
\bibitem{omega}     G. Belanger, F. Boudjema, A. Pukhov, A. Semenov, Comput. Phys. Commun. {\bf 149}, 103 (2002).
\bibitem{feynhiggs} M. Frank, S. Heinemeyer, W. Hollik, G. Weiglein, hep-ph/0212037
\bibitem{hdecay}    A. Djouadi, J. Kalinowski, M. Spira, Comput. Phys. Commun. {\bf 108}, 56 (1998).
\bibitem{suspect}   A. Djouadi,  J.-L. Kneur, G. Moultaka, hep-ph/0211331.
\bibitem{bsg}       H. Baer, C. Balazs, J. Ferrandis, X. Tata, \PRD64, 035004 (2001).
\bibitem{hagiwara}  K. Hagiwara, A. D. Martin, D. Nomura,  T. Teuber,  \PLB557, 69 (2003).
\bibitem{davier}    M. Davier, S. Eidelman, A. Hocker, Z. Zhang, Eur. Phys. J. C{\bf 27}, 497 (2003).
\bibitem{LHC}       H. Baer, A. Belyaev, T. Krupovnickas, A. Mustafayev, hep-ph/0403214.
\bibitem{su}        J. L. Feng, S. Su, F. Takayama, hep-ph/0404198; hep-ph/0404231.
\end{thebibliography}
\end{document}